\documentclass[usenatbib]{mn2e}

\usepackage{graphicx}
\usepackage{amssymb}
\newcommand{\link}[1]{{\tt#1}}

\newcommand{\aj}{AJ}
\newcommand{\apj}{ApJ}
\newcommand{\apjl}{ApJ}
\newcommand{\apjs}{ApS}
\newcommand{\aap}{A\&A}
\newcommand{\mnras}{MNRAS}
\newcommand{\nat}{Nature}

\newcommand{\atel}{ATel}%{The Astronomer's Telegram}
\newcommand{\iaucirc}{IAUC}%{International Astronomical Union Circulars}

\newcommand{\pasj}{PASJ}
\newcommand{\pasa}{PASA}
\newcommand{\apss}{Ap\&SS}

\newcommand{\araa}{ARA\&A}%

\newcommand{\japa}{JApA}%Journal of Astrophysics and Astronomy

\newcommand{\ascl}{Astrophysics Source Code Library}%Astrophysics Source Code Library

\newcommand{\arcdeg}{\ensuremath{^{\circ}}}

\newcommand{\nH}{\ensuremath{N_{{\rm H}}}}

\newcommand{\xtenineteen}{XTE\,J1908$+$094}

\newcommand{\swift}{\emph{Swift}}

\newcommand{\lmxb}{LMXB}
\newcommand{\lmxbs}{LMXBs}

\title
[Polarimetry as a probe of unresolved jets: \xtenineteen]
{Radio polarimetry as a probe of unresolved jets: the 2013 outburst of \xtenineteen}

\author[P.A.~Curran et al.]
{P.A.~Curran$^{1}$\thanks{e-mail: peter.curran@curtin.edu.au},
J.C.A.~Miller-Jones$^{1}$, 
A.P.~Rushton$^{2,3}$, %Oxford
D.D.~Pawar$^{4}$, %S'ton
G.E.~Anderson$^{2}$,\newauthor %Oxford
D.~Altamirano$^{3}$, %S'ton
H.A.~Krimm$^{5,6}$,
J.W.~Broderick$^{2}$, %Oxford
T.M.~Belloni$^{7}$,   %INAF
R.P.~Fender$^{2}$, \newauthor %Oxford
E.G.~K\"ording$^{8}$,  % Nijmegen
D.~Maitra$^{9}$, %Wheaton
S.~Markoff$^{10}$,  %API
S.~Migliari$^{11,12}$,  % Barcelona
C.~Rumsey$^{13}$, %Cambridge %AMI
M.P.~Rupen$^{14}$, \newauthor %BC, NRAO
D.M.~Russell$^{15}$, %NYU
T.D.~Russell$^{1}$,  %Curtin
C.~L.~Sarazin$^{16}$, %Virginia
G.R.~Sivakoff$^{17}$, %U Alberta
R.~Soria$^{1}$,\newauthor %Curtin
A.J.~Tetarenko$^{17}$,  % %U Alberta
D.~Titterington$^{13}$,  %Cambridge% AMI
V.~Tudose$^{18}$   %Bucharest
\\
$^{1}$International Centre for Radio Astronomy Research, Curtin University, GPO Box U1987, Perth, WA 6845, Australia\\ 
$^{2}$Astrophysics, Department of Physics, University of Oxford, Keble Road, Oxford OX1 3RH, UK \\ 
$^{3}$School of Physics and Astronomy, University of Southampton, Southampton, Hampshire, SO17\,1BJ, UK \\ 
$^{4}$Department of Physics, R. J. College, Ghatkopar (W), Mumbai 400086, India\\
$^{5}$Universities Space Research Association, Columbia, MD 21044, USA \\ %Krimm
$^{6}$NASA/Goddard Space Flight Center, Astrophysics Science Division, Code 661, Greenbelt, MD 20771, USA  \\ %Krimm
$^{7}$INAF -- Osservatorio Astronomico di Brera, Via E. Bianchi 46, I-23807, Merate (LC), Italy \\
$^{8}$Department of Astrophysics/IMAPP, Radboud University Nijmegen, PO Box 9010, NL-6500 GL Nijmegen, the Netherlands \\
$^{9}$Department of Physics and Astronomy, Wheaton College, Norton, MA 02766, USA \\
$^{10}$Astronomical Institute `Anton Pannekoek', University of Amsterdam, P.O. Box 94249, 1090 GE Amsterdam, the Netherlands \\
$^{11}$XMM-Newton Science Operations Centre, ESAC/ESA, PO Box 78, 28691 Villanueva de la Ca\~nada, Madrid, Spain\\
$^{12}$Department of Astronomy and Meteorology \& ICCUB, University of Barcelona, Mart\'i i Franqu\`es 1, 08028 Barcelona, Spain\\
$^{13}$Astrophysics Group, Cavendish Laboratory, 19 JJ\,Thomson Avenue, Cambridge CB3 0HE, UK\\
$^{14}$National Research Council, Herzberg Astronomy and Astrophysics, 717 White Lake Road, PO Box 248, Penticton, BC V2A 6J9, Canada\\
$^{15}$New York University Abu Dhabi, P.O. Box 129188, Abu Dhabi, United Arab Emirates \\
$^{16}$Department of Astronomy, University of Virginia, P.O. Box 400325, Charlottesville, VA 22904, USA \\
$^{17}$Department of Physics, University of Alberta, CCIS 4-181, Edmonton, AB T6G 2E1, Canada \\
$^{18}$Institute for Space Sciences, Atomistilor 409, P.O. Box MG-23, Bucharest-M\v{a}gurele RO-077125, Romania \\
}

\begin{document}

\date{Accepted 2015 June 02.  Received 2015 June 01; in original form 2015 March 03}

\pagerange{\pageref{firstpage}--\pageref{lastpage}} \pubyear{}

\maketitle

\label{firstpage}

%%**************************************************
%% Abstract
%%**************************************************

\begin{abstract}
\xtenineteen\ is an X-ray transient black hole candidate in
the Galactic plane that was observed in outburst in 2002 and 2013.
Here we present multi-frequency radio and X-ray data, including radio
polarimetry, spanning the entire period of the 2013 outburst. 
We find that the X-ray behaviour of \xtenineteen\ traces the standard
black hole hardness-intensity path, evolving from a hard state,
through a soft state, before returning to a hard state and quiescence.
Its radio behaviour is typical of a compact jet that becomes quenched
before discrete ejecta are launched during the late stages of X-ray
softening.
The radio and X-ray fluxes, as well as the light curve morphologies,
are consistent with those observed during the 2002 outburst of this
source.
The polarisation angle during the rise of the outburst infers a jet
orientation in agreement with resolved observations but also displays
a gradual drift, which we associate with observed changes in the
structure of the discrete ejecta.  We also observe an unexpected
90\arcdeg\ rotation of the polarisation angle associated with a second
component.
\end{abstract}


%%**************************************************
%% Section 1: Introduction
%%**************************************************

\section{Introduction}\label{section:introduction}

Relativistic jets are a standard feature of black holes in both
actively accreting (e.g.,
\citealt{Mirabel1998:Natur.392,Fender2006:csxs.book381}) and quiescent
(e.g., \citealt{Gallo2006:MNRAS.370}) low mass X-ray binary (\lmxb)
systems. The evolution of the bipolar jets during an outburst event is
dependent on the accretion rate onto the black hole
\citep{Fender2006:csxs.book381} and is closely coupled with the
observed X-ray ``states'' (see e.g.,
\citealt{mclintock2006:csxs157,belloni_2010LNP...794}).
During the {\it hard} X-ray state of a standard \lmxb\ outburst (when
the X-ray spectrum is dominated by power-law emission from the
optically-thin, geometrically-thick inner regions), radio emission
from compact, partially self-absorbed, flat-spectrum ($\alpha \sim 0$,
where $F_{\nu} \propto \nu^{\alpha}$) jets is often observed.
The radio emission of the compact jets increases in luminosity, from
quiescent levels, before being quenched by a factor of at least
$\sim$700 (e.g., \citealt{Coriat2011:MNRAS.414,Russell2011:ApJ.739})
at, or around, the X-ray peak of the outburst -- corresponding to the
transition to the {\it soft} X-ray state (where the X-ray spectrum is
dominated by a thermal blackbody component from the accretion disk).
During this transition to the soft state, defined by various classes
of {\it intermediate} X-ray states, the overall radio emission is
observed to become optically thin ($\alpha < 0$) and, occasionally, to
exhibit flares (though these may be easily missed due to low-cadence
radio observations). In fact, in a number of sources, this
optically-thin emission has been spatially resolved as discrete ejecta
(e.g.,
\citealt{Tingay1995:Natur.374,Mirabel1998:Natur.392,Miller-Jones2012:MNRAS.421}).
However, it is also worth noting that in some cases -- such as the 2010
outburst of  MAXI\,J1659$-$152, which was observed to
transition to a full soft state -- no evidence of discrete ejecta were
detected despite being well sampled at the time 
\citep{Paragi2013:MNRAS.432,vanderHorst2013:MNRAS.436}.
Finally, towards the end of the outburst, when the X-ray luminosity
decreases to approximately 2\% of the Eddington luminosity
\citep{Maccarone2003:A&A.409,Dunn2010:MNRAS.403,Kalemci2013:ApJ.779},
the compact jets are reestablished and seen to fade with X-ray
luminosity.

Optically-thin radio emission from the discrete ejecta can, in the
presence of an ordered magnetic field, be linearly polarised at a
level of up to $\approx 70\%$ \citep{Longair1994:hea2.book}. The
flat-spectrum compact jets can reach a fraction of this, depending on
the underlying distribution of synchrotron spectra.
However, the magnetic field is not necessarily ordered and a number of
mechanisms -- such as multiple, unresolved components that cancel each
other out, or spatially dependent Faraday rotation -- can suppress the
resulting net polarisation (see e.g.,
\citealt{Brocksopp2007:MNRAS.378} and references therein).
Most \lmxbs\ with published fractional polarisations have been
observed at much lower levels of $\lesssim$10\% (e.g.,
\citealt{Fender2003:Ap&SS.288} and references therein;
\citealt{Brocksopp2007:MNRAS.378}) with relatively  few having
fractional polarisations of $\approx$50\%
\citep{Brocksopp2013:MNRAS.432,Curran2014:MNRAS.437}.
When detected, polarisation can be used to infer properties of the
magnetic field and the surrounding medium (e.g.,
\citealt{Stirling2004:MNRAS.354,Miller-Jones2008:ApJ.682}), as well as
properties of the jet itself.  An alignment between the compact jet
axis and the intrinsic polarisation angle has now been observed in a
number of sources (e.g.,
\citealt{Corbel2000:A&A.359,Russell2014:MNRAS.438,Russell2015:MNRAS.450})
and, if common to all black hole X-ray \lmxbs, this allows us to infer
the orientation of unresolved compact jets.

\subsection{\xtenineteen}\label{sec:intro:source}

The transient X-ray source, \xtenineteen\ was first detected on
21 February 2002 (MJD 52326) by the Proportional Counter Array (PCA)
aboard the {\it Rossi X-ray Timing Explorer} (RXTE) satellite
\citep{Woods2002IAUC.7856}.
Very Large Array radio observations on 21 March 2002
\citep{Rupen2002IAUC.8029} refined the position in the Galactic plane
($l,b = 43.26\arcdeg, +0.43\arcdeg$), to sub-arcsecond
precision\footnote{All uncertainties in this paper are quoted and/or
  plotted at the $1\sigma$ confidence level.}, as RA, Dec =
19:08:53.077, +09:23:04.9 $\pm$ 0.1\arcsec (J2000).
Further Very Large Array observations identified discrete ejecta,
moving along the north-northwest
direction\footnote{\link{http://www.aoc.nrao.edu/$\sim$mrupen/XRT/X1908+094/\\x1908+094.shtml}\label{rupen}}.
Two possible near-infrared counterparts were identified consistent
with this position
\citep{Chaty2002IAUC.7897,Chaty2006MNRAS.365}\footnote{Note that these
  authors misquote the RA as 19:08:53.77, though have used the
  correct position when identifying the counterpart.}.
On the basis of X-ray timing and spectral properties, including the
detection of quasi-periodic oscillations (QPOs) in the power spectrum,
the source was classified as a black hole candidate that exhibited
both hard and soft states \citep{Gogus2004ApJ.609,intZand2002A&A.394}.
Furthermore, limited quasi-simultaneous radio (Very Large Array and
Westerbork Synthesis Radio Telescope, WSRT) and X-ray (RXTE)
observations were found to be consistent with the expected hard state
correlation of black holes (e.g., \citealt{Gallo2003:MNRAS.344}) by
\cite{Jonker2004:MNRAS.351}.

\xtenineteen\ was next detected in outburst on 26 October 2013 (MJD
56591) by the Burst Alert Telescope (BAT) on the \swift\ satellite
\citep{Krimm2013ATel.5523} and further detected by both the
\swift\ X-ray Telescope (XRT) and the Monitor of All-sky X-ray Image
(MAXI) instrument aboard the International Space Station
\citep{Krimm2013ATel.5529,Negoro2013ATel.5549}. A radio counterpart
was detected at frequencies from 5 to 22 GHz at the Karl G. Jansky
Very Large Array (VLA; \citealt{Miller-Jones2013ATel.5530}), the
Arcminute Microkelvin Imager Large Array (AMI-LA;
\citealt{Rushton2013ATel.5532}), the Australia Telescope Compact Array
(ATCA; \citealt{Coriat2013ATel.5575}), the Very Long Baseline Array
(VLBA) and the European VLBI Network (EVN;
\citealt{Rushton2014}). 
\cite{Rushton2014} also confirm the 2002 direction of jet motion
along the north-northwest direction as well as identifying an
expanding Southern component that fades to background levels and is
superseded by a Northern component on MJD 56607.
Here we present our VLA, Low-Frequency Array (LOFAR) and AMI-LA
observations combined with X-ray monitoring of the 2013 outburst of
\xtenineteen.
In section \ref{section:observations} we introduce the observations
and reduction methods, while in section \ref{section:discussion} we
discuss the results of our photometric and polarimetric analyses of
the data and discuss their physical implications for the system. We
summarise our findings in section \ref{section:conclusion}.

%%**************************************************
%% Section 2: Method
%%**************************************************

\section{Observations \& Analysis}\label{section:observations}

\subsection{Radio data}\label{section:radio}

\subsubsection{VLA}\label{section:vla}

\xtenineteen\ was observed by the VLA from 29 October to 22 November
2013 (15 epochs in B configuration) and again on 11 and 22 March 2014
in A configuration (see Table \ref{table:fluxes}).
Observations were primarily made at central frequencies of 5.25 and
7.45\,GHz; each baseband was comprised of 8 spectral windows with
sixty-four 2\,MHz channels each, giving a bandwidth of 1.024\,GHz per
baseband. Further observations were obtained in the 18--26\,GHz band
(central frequency of 22\,GHz) using eight 1.024\,GHz basebands, as
above.
Additionally, on 3 epochs observations were made in the 1--2\,GHz
band, which has 16 spectral windows with sixty-four 1\,MHz channels
each. The 1--2\,GHz band is heavily affected by radio frequency interference
(RFI) that reduced the usable spectral windows to between 5 and 6
($\approx$320\,MHz), spread over the band, which was treated as a
whole. This led to a total bandwidth of $\approx$500\,MHz with an
effective frequency of 1.6\,GHz.

Flagging, calibration and imaging of the data followed standard
procedures and were carried out within the Common Astronomy Software
Application (CASA 4.2.1) package \citep{McMullin2007:ASPC376}.
The primary calibrator used as the bandpass and polarisation angle
calibrator, and to set the amplitude scale at all frequencies was
either 3C48 (J0137+331) or 3C286 (J1331+3030), depending
on the local sidereal time of the observations.
J1824+1044 was used as the secondary (phase) calibrator at 1.6\,GHz,
J1922+1530 was used at 5.25 and 7.45 GHz and J1856+0610 was used at
22\,GHz.
The polarisation leakage calibrator at 5.25 and 7.45 GHz was J2355+4950,
except on 11 March 2014 when J1407+2827 was used.

Images were generated with robust=0 weighting and phase
self-calibration was performed on a per-band basis, or for 5.25 and 7.45
GHz on a per-baseband or per-spectral window basis depending on the
signal to noise ratio (though we only tabulate results at the central
frequencies of the basebands).
The flux densities of the source were measured by fitting a point
source in the image plane (Stokes $I$), and, as is usual for VLA data,
systematic errors of 1\% should be added at frequencies $<$10\,GHz and
3\% at 22\,GHz. At 5.25 and 7.45 GHz, Stokes $Q$ and $U$ flux densities
were measured at peak (Table\,\ref{table:fluxes}).

\subsubsection{AMI-LA}\label{section:ami}

AMI-LA observed \xtenineteen\ from 30 October 2013 to 22 March 2014,
with a cadence of between 1--4 days. Each epoch lasted approximately 4
hours, centred around the local sidereal time of the peak elevation of
the source.  AMI-LA consists of $8\times12.8$\,m antennas with
baselines ranging from 18--110\,m, located at the Mullard Radio
Astronomy Observatory in the UK \citep[AMI
  Consortium:][]{Zwart2008:MNRAS.391}.  AMI-LA has an operational
range of 13.9--17.5\,GHz when using frequency channels 3--7, each with
a bandwidth of 0.72 GHz (channels 2, 3 and 8 are disregarded due to
their current susceptibility to interference). This gives a total
useable bandwidth of $\sim$3.6\,GHz, centred at 15.7\,GHz.

Initial data reduction was performed with the \texttt{Python}
\texttt{drive-ami} pipeline \citep{Staley2013:MNRAS.428}.
\texttt{drive-ami} utilises the \texttt{REDUCE} software package,
which is designed to take the raw AMI-LA data and automatically flag
for interference, shadowing, and hardware errors, conduct phase and
amplitude calibrations, and Fourier transforms the data into
\textit{uv}-FITS format \citep{Perrott2013:MNRAS.429}.  Short,
interleaved observations of J1856+0610 were used for phase
calibration. Imaging was conducted using the \texttt{chimenea} data
reduction pipeline that has been specifically designed to deal with
multi-epoch radio observations of transients
\citep{Staley2015:ascl1502}. The resulting peak AMI-LA flux densities
of \xtenineteen\ (Table\,\ref{table:fluxes}) were measured using
\texttt{MIRIAD} \citep{Sault1995:ASPC.77} and systematic errors of 5\%
\citep{Perrott2013:MNRAS.429} should be added in quadrature. Note that
the flux errors may be underestimated due to uncleanable artefacts
resulting from the nearby off-axis radio-bright source that may be
associated with the star forming region W49A
\citep{Webster:1971AJ.76}. For further details on the reduction and
analysis performed on the AMI observations see
\cite{Anderson2014:MNRAS.440}.

\begin{figure} 
  \centering 
% \resizebox{16cm}{!}{\includegraphics[angle=0]{fig.combined_radio.ps} }
  \resizebox{\hsize}{!}{\includegraphics[angle=0]{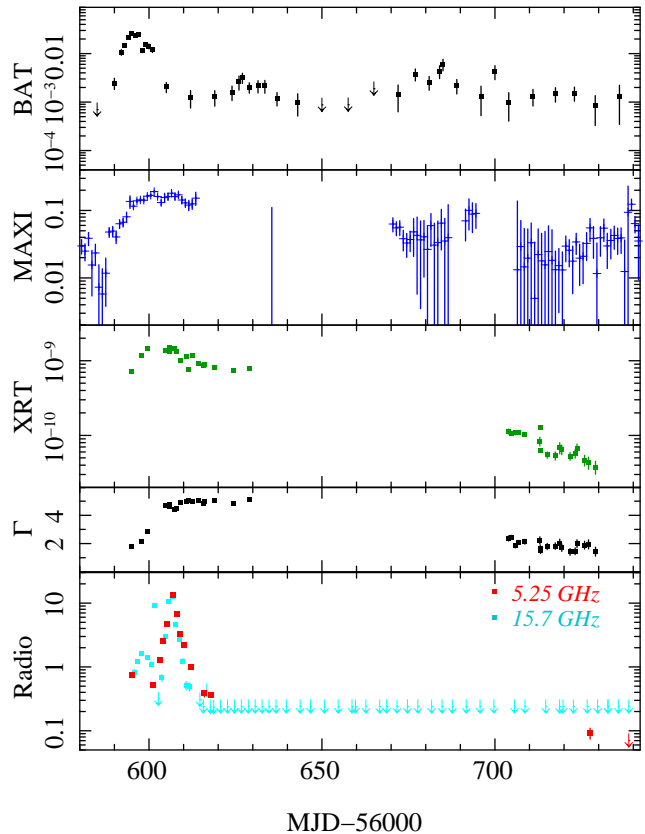} }
  \caption{Hard \swift/BAT 15--50\,keV, counts/second) and soft (MAXI
    2--20\,keV, counts/second$^{\ref{maxi}}$; \swift/XRT 3--9\,keV,
    ergs\,cm$^{-2}$\,s$^{-1}$) X-ray light curves, X-ray photon index
    ($\Gamma$), and radio (5.25\,GHz \& 15.7\,GHz, mJy) light
    curves. Due to the source's proximity to the bright, persistent
    source GRS\,1915+105, MAXI count-rates may suffer from
    contamination, especially at low luminosities.
}
 \label{fig:broad_lc} 
\end{figure}

\begin{figure*} 
  \centering 
  \resizebox{15cm}{!}{\includegraphics[angle=0]{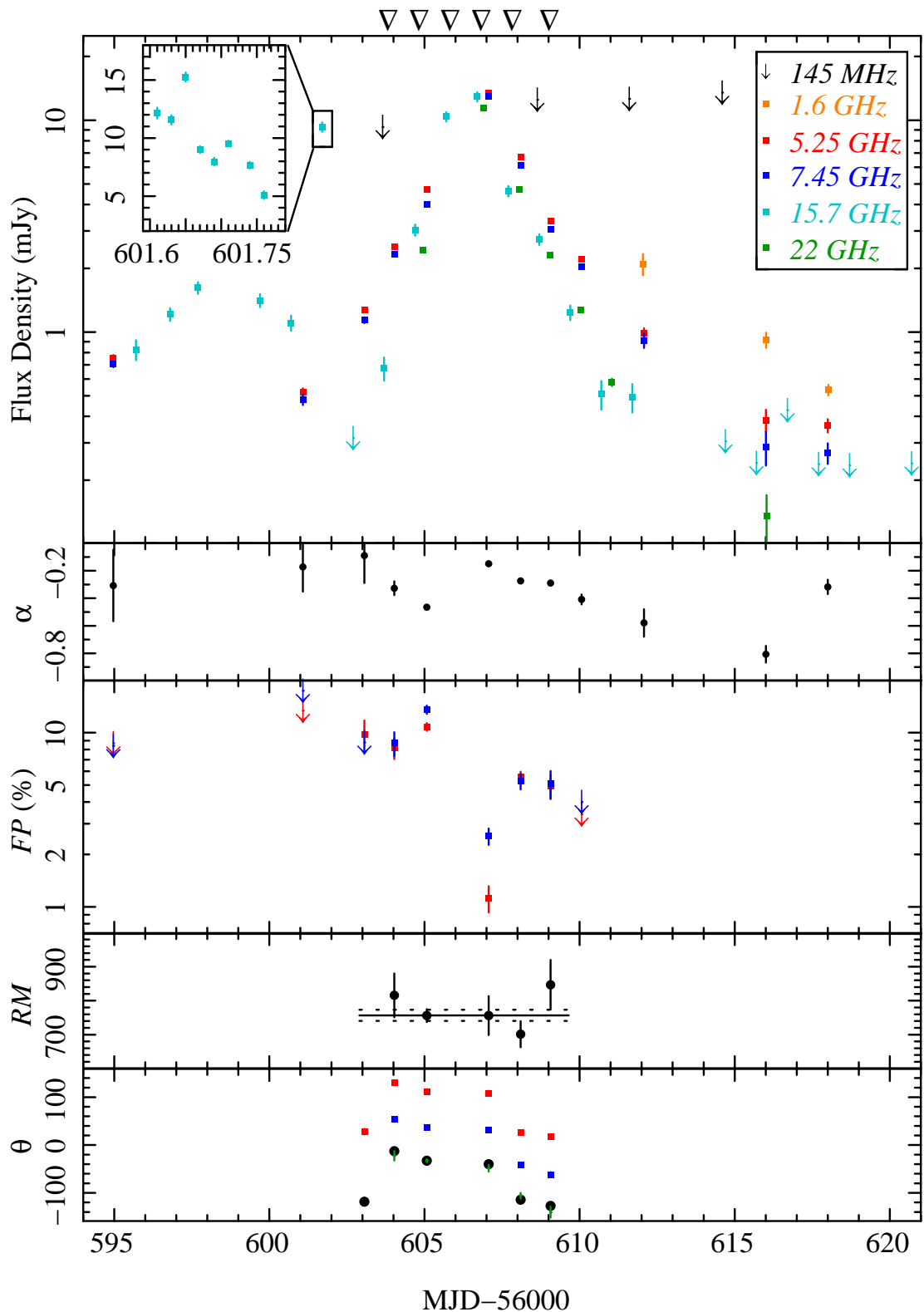} }
  %\resizebox{\hsize}{!}{\includegraphics[angle=0]{fig.Radio_curves.ps} }
  \caption{Radio light curves, spectral, and polarisation parameters
    of the source during outburst.
The upper panel shows the flux densities at each of the 6 radio
frequencies observed (legend is common to all panels), 
while the inlay shows the
time-resolved 15.7\,GHz data from MJD 56601.7.
The second panel shows the radio spectral index, $\alpha$ (where
$F_{\nu} \propto \nu^{\alpha}$), while the third panel shows the
fractional polarisations or limits, {\it FP}, at 2 frequencies (5.25
and 7.45 GHz; note that for clarity we only plot constraining limits,
i.e., $\leq20$\%.).
The fourth panel shows the rotation measure, $RM$, from independent
(data points) and simultaneous (solid line with $1\sigma$ errors as
dashed lines) fitting of the epochs. 
Panel five shows the observed polarisation angles (blue and red) and
the intrinsic electric vector position angle (EVPA), $\theta$, derived
from the independent (green) and simultaneous (black) fits of those PAs.
The inverted triangles along the top axis represent the epochs of the
VLBA observations reported in \protect\cite{Rushton2014}.  
}
 \label{fig:Radio_curves} 
\end{figure*}

\subsubsection{LOFAR}\label{section:lofar}

LOFAR \citep{vanHaarlem2013:A&A556} observed \xtenineteen\ on four
epochs over the period 7--18 November 2013. Observations were obtained
with the high-band antennas (HBA) from 115-189\,MHz, using 380
sub-bands, each with a bandwidth of 195.3\,kHz.
Each observation consisted of 20 minute snapshots of the field,
interleaved with 1.5 minute scans of the primary calibrator, 3C380,
for a total of 160 minutes on source in all epochs, except the second
observation where the total was 80 minutes. The runs were mainly
carried out during the daytime, and included transit; the minimum
source elevation across all four observations was about $39^{\circ}$.
Flagging, calibration and imaging followed standard methods
\citep[e.g.][]{Heald2011:JApA32,Offringa2012A&A.539,vanHaarlem2013:A&A556}
while amplitude calibration was based on the 3C380 model from
\citet{Scaife2012MNRAS.423}. We performed phase-only calibration on
the target field using a local sky model determined by
cross-correlating the 74 MHz VLA Low-Frequency Sky Survey
\citep[VLSS;][]{Cohen2007AJ.134} with the 1.4 GHz NRAO VLA Sky Survey
\citep[NVSS;][]{Condon1998AJ.115}.
Sub-bands were combined into six, approximately evenly-split bands and
an image was generated for each band with the task {\sc awimager}
\citep{Tasse2013:A&A553}, using robust $=0$ weighting.  These were
then convolved to a common resolution and averaged, using
inverse-variance weighting, to produce a final image. The projected
baseline range was restricted to 0.1--6 k$\lambda$ (maximum baseline
approximately 12 km in the centre of the band). Although we used the
full Dutch array in our observations, at the time of writing,
long-baseline imaging was not routine with simple, computationally
inexpensive calibration procedures such as the one we have adopted
here. The minimum baseline cutoff was empirically determined so as to
reduce the confusion limit due to extended emission along the Galactic
Plane.
The effective frequency of each image is about 145 MHz, and the
average angular resolution is 55\arcsec $\times$ 33\arcsec (with a
beam position angle of $\approx$30\arcdeg). No source was detected at
the position of \xtenineteen, on any epoch, above the flux limits
given in Table\,\ref{table:fluxes}.

\subsection{X-ray data}\label{section:xray}

We analysed 38 \swift/XRT Windowed Timing (WT) observations of
\xtenineteen\ obtained from 29 October to 3 December 2013 and from 15
February to 17 March 2014 (Table \ref{table:fluxes_X}). Between these
dates XRT observations were not possible due to the position of the
Sun (i.e. source was Sun-constrained).
Due to the known issue of low-energy spectral residuals in WT data for
heavily absorbed
sources\footnote{\link{http://www.swift.ac.uk/analysis/xrt/digest\_cal.php\#abs}}
we extracted source and background spectra in the 1.5--10\,keV range
only (grades 0 -- 2), using suitable extraction regions of radius
$\sim$75\arcsec.
Ancillary response files were generated using the \textsc{ftool} {\tt
  xrtmkarf} and the response matrix available at the time of
observations. 
Spectra were rebinned to have a minimum of 20 counts per bin (so that
$\chi^2$ statistics are valid) and analysed within \textsc{xspec}
(version: 12.8.0), though 2 late epochs did not have enough counts to
produce useful spectra (\swift\ Obs IDs 00033014037 \& 039).

Due to low count rates and the relatively high low-energy cutoff at
1.5\,keV, we are unable to consistently distinguish between the
physically-motivated absorbed multi-coloured black-body plus power-law
model ({\tt tbabs*(diskbb+pow)}) and the single absorbed power-law
model ({\tt tbabs*powerlaw}). Therefore we opt to fit a
phenomenological power-law model ({\tt tbabs*powerlaw}), purely to
calculate the flux from 3.0 to 9.0 keV (Table \ref{table:fluxes_X},
Figure\,\ref{fig:broad_lc}) and, via the photon index, to indicate the
X-ray state of the system.
WT mode data are unaffected by pile-up. 
The absorption, \nH, was in the range 1.5--3.0 $\times
10^{22}$\,cm$^{-2}$ (consistent with the value found by
\citealt{intZand2002A&A.394}) and the reduced $\chi^2$ varied from
0.78 to 1.96, with a mean value of 1.29.

Hard X-ray (15--50\,keV) light curves (counts/sec) were obtained by
the \swift/BAT transient
monitor\footnote{\link{http://swift.gsfc.nasa.gov/docs/swift/results/transients}}
\citep{Krimm2013:ApJS.209}. The data points are based on daily average
results from the monitor, which have been binned so that each point
either has a significance of $>$3 sigma, or covers a period of seven
days (Figure\,\ref{fig:broad_lc}).
For comparison MAXI \citep{Matsuoka2009:PASJ.61} light curves were
downloaded\footnote{\link{http://maxi.riken.jp}\label{maxi}} but due to the
source's proximity to the bright, persistent source GRS\,1915+105,
these may suffer from contamination, especially at low luminosities
\citep{Negoro2013ATel.5549}.

%%**************************************************
%%**************************************************
%%**************************************************

\subsection{Radio spectral indices}\label{section:indices}

We obtain the spectral index, $\alpha$ (where $F_{\nu} \propto
\nu^{\alpha}$), of the radio spectrum at each epoch with multiple
observed frequencies by fitting a power-law to the derived flux
densities, $F_{\nu}$, against frequency, $\nu$.
Due to the short time-scale variability of this source
(section \ref{section:discribe}) we use only data which are
simultaneous, i.e., the VLA data. 
The flux densities over the entire 1.6 and 22 GHz bands were used
along with the flux densities at 5.25 and 7.45 GHz on a per-baseband or,
for the brighter epochs, a per-spectral window basis. We obtain
similar values for the spectral indices using both the per-baseband
and per-spectral window flux densities.
All epochs were well fit by a single power law with no need for
additional components.

\subsection{Polarisation}\label{section:polarisation}

\begin{figure} 
 \centering 
 \resizebox{\hsize}{!}{\includegraphics[angle=0]{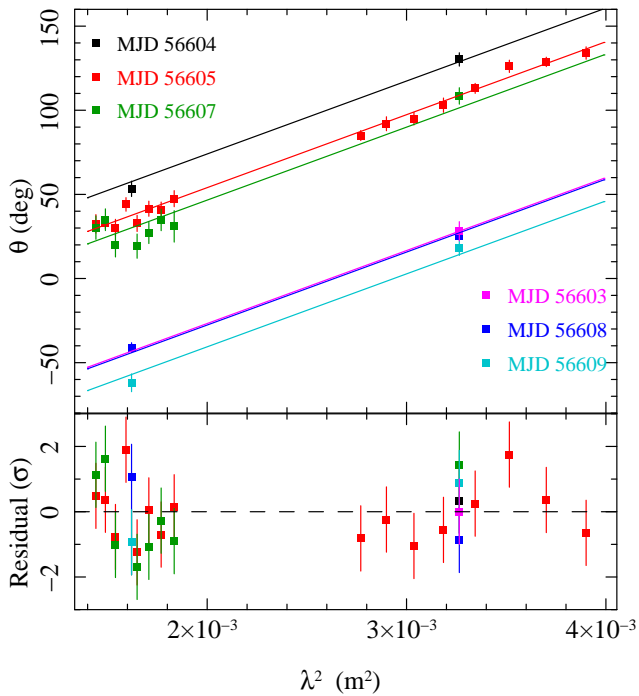} }
  \caption{Plot of observed polarisation angle versus wavelength, and
    simultaneous fit for the rotation measure, for the 6 epochs with
    observed polarisations. The lower panel shows the residuals (in
    units of $\sigma$) from the simultaneous fit.}
 \label{fig:Pangle} 
\end{figure}

We derived the polarisation parameters at 5.25 and 7.45 GHz from the
measured flux densities of the Stokes $I$, $Q$ and $U$ images
(Table\,\ref{table:fluxes}):
linear polarisation, $LP = \sqrt{Q^2 + U^2}$;
fractional polarisation, $FP = 100~LP/I$;
and polarisation angle, $PA = 0.5 \arctan(U/Q)$, which is degenerate
such that derived angles may be offset by an integer multiple of $\pm
180$\arcdeg\ from the true value.
In the case of non-detections, the upper limit on polarimetry,
$P_{{\rm limit}}$, is calculated as 3 times the RMS noise of the
linear polarisation over the field. 
The derived polarimetric parameters at 5.25 and 7.45 GHz are plotted in
Figure\,\ref{fig:Radio_curves}.

Faraday rotation in the local or interstellar medium causes a rotation
of the polarisation vectors at wavelength, $\lambda$, such that the
intrinsic electric vector position angle (EVPA) of the source is
related to the observed polarisation angle, {\it PA}, by ${\rm EVPA} =
PA - RM\lambda^2$ where $RM$ is the rotation measure
(e.g., \citealt{Saikia-Salter1988:ARA&A.26}).
Given observed polarisation angles, the rotation measures and EVPAs
(plotted in Figure\,\ref{fig:Radio_curves}) are derived from a linear
fit of $PA$ versus $\lambda^2$ for epochs where we were able to derive
polarisation angles at multiple wavelengths. Independent fitting of
each epoch returned consistent rotation measures, with a weighted
average of $\approx 775$\,rad\,m$^{-2}$. Applying a modified version
of the simultaneous fit described by \cite{Curran2007A&A.467} we find
a common rotation measure of $757 \pm16$\,rad\,m$^{-2}$ and EVPAs that
agree with those of the independent fits ($\chi^{2}_{\nu}=1.2$;
Figure\,\ref{fig:Pangle}).

%%**************************************************
%%**************************************************
%%**************************************************

\section{Results and discussion}\label{section:discussion}

\subsection{Light curves \& spectral indices}\label{section:discribe}

The broadband light curves (Figure\,\ref{fig:broad_lc}) of the
monitoring observations start to rise at MJD$\sim$56588--56590.
Initially, these peak in the hard X-rays (15--50\,keV), at
MJD$\sim$56595, before fading by MJD 56610. 
The softer X-rays (2--20\,keV) have a more gradual rise and decline,
peaking at MJD$\sim$56601. By MJD$\sim$56675, after a period of the
source being Sun-constrained, they had returned to the RMS noise level
of 0.0021 photons\,cm$^{-2}$\,s$^{-1}$ (excluding outburst), though
there may be hints of activity at later times.
The 3--9\,keV X-rays peak at MJD$\sim$56606 and decay to quiescence at
MJD$>$56729, though there is a wide range of MJDs that are not sampled
when the source was unobservable due to the position of the Sun.

The 5.25 and 7.45 GHz radio light curves (see also
Figure\,\ref{fig:Radio_curves}) peak at a similar time to the
3--9\,keV X-rays, at MJD 56607, but have shorter rise and decay times
of only $\sim$5 days. The higher-cadence 15.7\,GHz observations reveal
an sharp flare, peaking at MJD$\sim$56601.7, at the same time as the
peak in the 2--20\,keV X-rays. This flare rises by almost an order of
magnitude in a day and falls by a factor of at least 26 on the same
time-scale. A time-resolved analysis of this data point (see inlay of
Figure\,\ref{fig:Radio_curves}) reveals significant variability over
the 4-hour observation, with an approximate decay consistent with the
non-detection a day later.

During the X-ray flux rise the observed photon index increased from
$\Gamma \sim 2$ to $\sim 5$, where it remained until the
Sun-constrained period. 
Noting that we were unable to fit a physically motivated model to the
\swift\ spectra (section \ref{section:xray}), we define the soft state
as the period when the photon index was $\sim$5; this definition is
consistent both with the reported transition to a soft state at
MJD$\sim$56598 \citep{Negoro2013ATel.5549} and with the significant
drop of hard (15--50\,keV) X-ray photons at that time.
When X-ray observations restarted, on MJD 56703, the photon index of
our phenomenological fit was at a much lower value ($\Gamma <2.5$),
suggesting that the source had returned to a hard state; though the
still-decreasing photon index indicates that the source was still
hardening through to MJD 56730.

The observed radio spectra have indices, $\alpha$, ranging from $-0.1$
to $-0.8$ (Figure\,\ref{fig:Radio_curves}). Initially consistent with
being optically thick ($\alpha \approx 0$), the indices steepen during
the rise to maximum, at which point they flatten before again
steepening with decreasing flux. The 3 initial spectral indices are
ambiguous but the non-zero values could be consistent with the X-ray
source transitioning to a soft state at MJD$\sim$56598
\citep{Negoro2013ATel.5549}.
On 7 epochs, we have low-frequency (145\,MHz or 1.6\,GHz) flux limits
or detections. On the 3 epochs where we have 1.6\,GHz fluxes, the
spectra are consistent with a single (optically thin) spectral index
at each epoch, over the range of all observed frequencies
(1.6--22\,GHz). We have flux limits at 145\,MHz on 4 epochs but only
the 3$\sigma$ limit on MJD $\sim$56609 is constraining, at a value of
$\leq$12.5\,mJy.  Interpolating the 5.25, 7.45 and 22 GHz to the same
epoch as our 145\,MHz observations, and assuming a single power-law
spectral index, indicates that \xtenineteen\ would have been detected
at a level of $\approx$$13 \pm 2$\,mJy. While this is not inconsistent
with the non-detection, closer inspection of the 145\,MHz image
reveals no candidate at a lower significance (formal value at position
is $-$5\,mJy), suggesting a possible break in the spectrum. The higher
frequencies have a spectral index of $\alpha\approx-0.26\pm0.03$,
consistent with optically-thin emission from discrete ejecta, which
suggests that the low-frequency emission at this time could be
self-absorbed ($\alpha\approx+2.5$), as previously observed at
$\sim$1\,GHz for a discrete ejection of Cygnus\,X-3
\citep{Miller-Jones2004:ApJ600}.
Future low-frequency observations from e.g., LOFAR or the Murchison
Widefield Array (MWA), will allow us to constrain the value and study
the evolution of the spectral breaks over a wider range of frequencies
than is currently possible and hence, in the case of compact jets,
constrain the length of the jet.

\subsubsection{Hardness--Intensity}

\begin{figure}
  \centering 
  \resizebox{\hsize}{!}{\includegraphics[angle=0]{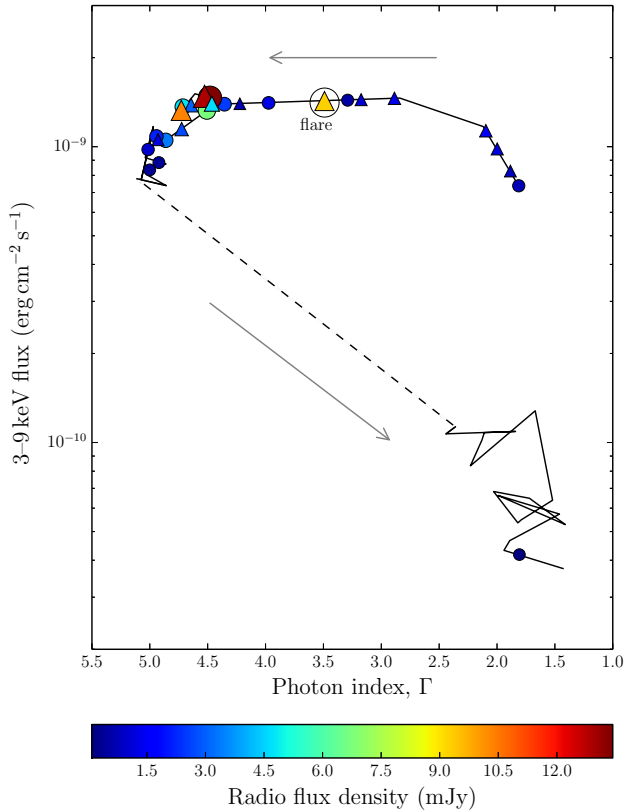} }
  %\vspace{0mm} 
  \caption{Proxy \swift\ Hardness--Intensity diagram (photon index
    versus flux) of the outburst, superimposed with 5.25 (circle) and
    15.7 GHz (triangle) radio flux densities, with arrows representing
    temporal direction. The dashed line represents the period when
    X-ray observations were unavailable due to the source being
    Sun-constrained.  The radio flux peaked in the soft (or possibly,
    late-intermediate) state while the sharp 15.7\,GHz radio flare at
    MJD 56602 (circled) occurred during the intermediate state.
}
  \label{fig:HID} 
\end{figure}

We plot the X-ray photon index versus the 3-9\,keV flux as a proxy for
the standard hardness-intensity diagram (HID in
Figure\,\ref{fig:HID}). This clearly shows that the source started in
a hard state before quickly softening at constant flux before dimming
at its softest. By the time the source emerged from being
Sun-constrained (see section \ref{section:xray}), it had reverted to
the hard state at a low luminosity. The source therefore followed the
standard trajectory
(e.g. \citealt{Homan2001:ApJS.132,homan_2005ApSS...300,belloni_2010LNP...794})
through our proxy-HID.

Though the X-ray spectra do not allow us to unambiguously infer the
states, the superimposed radio flux densities peak in either the
late-intermediate or in the soft state, while the sharp radio flare,
at MJD 56602, occurred during the intermediate state. Both of these
radio flux maxima are consistent with the region associated with
bright emission from a recent discrete ejection event \citep{Fender2006:csxs.book381}. 
The radio detection in the late, hard X-ray state indicates that the
compact jet had been reestablished by MJD 56727, though due to limited
VLA sampling and the AMI-LA flux limits it is unclear when it was
reactivated. The 5.25--7.45\,GHz spectral index at the time ($\alpha = 0.0
\pm 0.9$, Table\,\ref{table:fluxes}) is unconstraining.

\subsubsection{Comparison to 2002 outburst}

The 2013 outburst of \xtenineteen\ lasted about 150 days (as observed
by XRT), with a soft state of between 25 and 100 days. The total
observed outburst is therefore significantly shorter than the 2002
(RXTE-observed) outburst \citep{Gogus2004ApJ.609}, which lasted
$\sim$400 days, staying active into 2003; though the duration of the
soft state, lasting 60 days, could be comparable. Our rise time of
$\sim$15--20\,days is significantly quicker than the RXTE observed
rise times in 2002 of 50--80\,days, though instrumental and
energy-range differences will have a major effect. Overall however,
the X-ray morphology displays no obvious differences.
We observed a peak X-ray (3--9\,keV) flux of $\approx$1.8 $\times
10^{-9}$ ergs\,cm$^{-2}$\,s$^{-1}$ on MJD 56609. In 2002 the peak
observed X-ray (2.5--25\,keV) flux was 2.8 $\times 10^{-9}$
ergs\,cm$^{-2}$\,s$^{-1}$, on MJD 52355 \citep{Gogus2004ApJ.609},
which corresponds to $\approx$1.1 $\times 10^{-9}$
ergs\,cm$^{-2}$\,s$^{-1}$ in the 3--9\,keV range, assuming a power-law
model.

It is difficult to directly compare the peak radio flux due to
poor temporal sampling over the 2002 outburst. As our observations
show (Figure\,\ref{fig:Radio_curves}), without regular radio
observations many radio flares may go undetected, i.e., the sharp
radio flare at MJD 56602 was only detected in our 15.7\,GHz observations
and there was no evidence of it in our less frequent 5.25\,GHz data.
In 2002$^{\ref{rupen}}$ the highest reported flux density was 1.8\,mJy
at 5.25\,GHz, $\lesssim$10 times dimmer than our maximum flux density
(13\,mJy at 5.25\,GHz) but consistent with most of our measurements,
which are at the mJy level. Likewise, when we plot X-ray versus radio
luminosity (Figure\,\ref{fig:Luminosity}; see section
\ref{sect:energetics}) there is no difference between data from the
two outbursts.
We therefore find that there is no significant difference between
either the X-ray or radio flux behaviour of the two outbursts, despite
the fact that the resolved, discrete ejecta were observed
$\sim$0.5\arcsec\ apart in 2002$^{\ref{rupen}}$ and a factor of 10
less in 2013 \citep{Rushton2014}.

\subsection{Polarisation}

The radio emission at 5.25 and 7.45 GHz exhibits modest but variable
levels of polarisation of 1--14\% during the peak of the outburst
(Figure\,\ref{fig:Radio_curves}). At other epochs, we place upper
limits on the fractional polarisation but at MJDs $>$56611 these
limits are over 20\% and are not constraining. The fractional
polarisation is at a minimum at the peak flux, on MJD 56607, though
the absolute polarised flux density remains relatively constant
throughout at 0.1-0.5\,mJy so the minimum fractional polarisation may
be caused by the increase in total intensity.
The derived rotation measure of $757 \pm16$\,rad\,m$^{-2}$ over all
epochs is marginally higher than the RMs previously measured in this
region, though those exhibit significant variation with $-600 \lesssim
RM \lesssim 600$ rad\,m$^{-2}$
\citep{Taylor2009ApJ.702,Kronberg2011PASA.28}.
The observed polarisation angles and inferred EVPA exhibit a clear
$\approx90$\arcdeg\ rotation from $-120$\arcdeg\ on MJD 56603 to an
average (though slightly decreasing) value of $-30$\arcdeg\ between
MJDs 56604 and 56607; they are then observed to return to their
original values on MJD 56608.

\subsubsection{Electric vector position angle \& magnetic field}\label{sect:EVPA}

\begin{figure}
  \centering 
  \resizebox{\hsize}{!}{\includegraphics[angle=0]{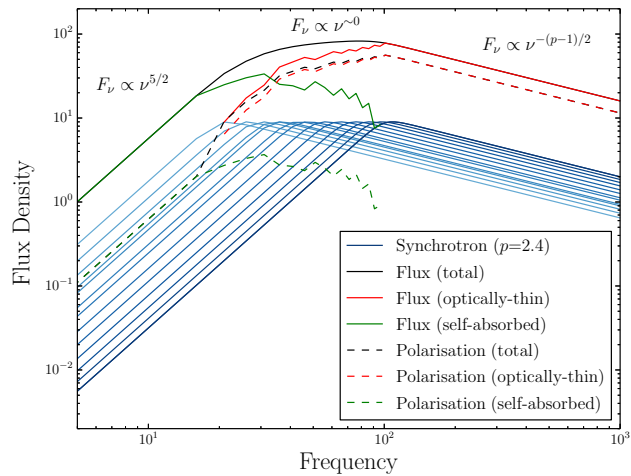} }
  \caption{Schematic representation of the flux and polarisation from
    a compact, partially self-absorbed jet demonstrating that the
    polarisation of the flat spectrum is dominated by optically-thin
    regions. The underlying distribution of the electron populations
    producing the individual synchrotron spectra is arbitrary, as are
    the flux and frequency units.}
 \label{fig:synchrotron} 
\end{figure}

Before we discuss \xtenineteen, let us first consider the expected
theoretical relationship between EVPA and magnetic field for both
compact, partially self-absorbed, flat-spectrum jets and
optically-thin discrete ejecta.
The radio emission originating from an optically-thin discrete
ejection is that of a single synchrotron spectra while the emission
from a compact, partially self-absorbed jet is a superposition of
multiple synchrotron spectra (e.g.,
\citealt{Blandford1979:ApJ232,Hjellming1988:ApJ.328}), each with both
optically-thin and self-absorbed regions. Since $F_{\nu} \propto
\nu^{+2.5}$ at self-absorbed frequencies and $F_{\nu} \propto
\nu^{-(p-1)/2} \approx \nu^{-0.7}$ at optically-thin frequencies
(assuming a typical value of the electron energy distribution index,
$p=2.4$, e.g., \citealt{Curran2010:ApJ.716}) the combined, flat
spectrum (Figure\,\ref{fig:synchrotron}) will be dominated by emission
from optically-thin regions of the individual synchrotron spectra
(e.g., \citealt{Zdziarski2014:MNRAS.442}).
Assuming a uniform magnetic field, optically-thin synchrotron emission
is expected to have a maximum fractional polarisation, $FP \lesssim
100(p+1)/(p+7/3) \approx 72\%$ while self-absorbed synchrotron
emission, on the other hand, has $FP \lesssim 300/(6p+13) \approx
11\%$ \citep{Longair1994:hea2.book}.  The maximum fractional
polarisation is therefore expected to be $FP \lesssim 72\%$ for
optically thin ejecta and a fraction of this, depending on the
underlying electron populations, for compact, partially
self-absorbed jets (Figure\,\ref{fig:synchrotron}).
Likewise, while the EVPA of self-absorbed synchrotron emission is
expected to be aligned parallel to the underlying magnetic field, the
observed EVPA of a partially self-absorbed source will be dominated by
optically-thin emission, which is aligned perpendicular to the
magnetic field (e.g.,
\citealt{Ginzburg1969:ARA&A.7,Longair1994:hea2.book}).  Therefore, the
EVPAs of both compact jets and discrete ejecta should be aligned
perpendicularly to their magnetic fields.

% South
Between MJDs 56604 and 5660, the observed EVPAs of \xtenineteen\ of
$\approx -30$\arcdeg\ originate from a magnetic field perpendicular to
the jet/ejection axis (north-northwest direction) observed in 2013 by
both the VLBA and EVN, and, in 2002, by the Very Large Array.
These epochs coincide with the period when the VLBA clearly detected a
dominant component moving to the south \citep{Rushton2014} . In the
simplest geometry of discrete ejecta, the dominant magnetic field is
caused by shock compression \citep{Laing1980:MNRAS.193} and is
parallel to the shock front (i.e., perpendicular to the jet axis). Our
observations are hence consistent with such a strong forward shock in
the southern component.
It is also interesting to note that the EVPA decreases and deviates
from the jet/ejection axis as the VLBA-observed Southern component
becomes spatially larger. This may indicate that magnetic field
compression due to a forward shock becomes weaker as the ejection
expands, allowing a differently orientated magnetic field to dominate
the polarised emission.

%North

The clear $\sim$90\arcdeg\ rotation of the EVPA between MJDs 56607 and
56608, just after the peak of the flux, coincides with the fading of
the southern component and the dominance of a new component moving to
the north (\citeauthor{Rushton2014}). Clearly, this discrete ejection
does not follow the simple geometry described above but a more complex
geometry due to, e.g., a complex magnetic field, lateral expansion,
velocity shear, etc. (see \citealt{Curran2014:MNRAS.437} and
references therein).
Without spatially-resolved polarimetry, as has been obtained for both
\lmxbs\ (e.g., \citealt{Miller-Jones2008:ApJ.682}) and, more commonly,
AGN (e.g.,
\citealt{Lister2005:AJ.130,Gomez2008:ApJ.681,Homan2009:ApJ.696}), it
is not possible to determine which of these, if any, explain the
non-standard magnetic field geometry.

Our initial EVPA, on MJD 56603, of $\approx -120$\arcdeg\ originates
from a magnetic field aligned in the north-northwest direction, along
the jet/ejection axis. It is unclear which of the two components, if
either, this originates from since there is no resolved imaging at
this time but it is unlikely that we are observing the compact
jet. The X-ray photon index suggests that the source is already in a
soft state and, while the radio data at this epoch are consistent with
either a flat or an optically thin spectrum, the sharp radio flare,
observed at 15.7\,GHz on MJD$\sim$56602, is indicative of a transition
to the soft state \citep{Fender2006:csxs.book381}.

Variable EVPAs, or \lq rotator events', similar to the
$\sim$90\arcdeg\ rotations observed here have previously been observed
in the \lmxbs\ GRO\,J1655$-$40 \citep{Hannikainen2000:ApJ.540} and
GRS\,1915+105 \citep{Fender2002:MNRAS.336}, and in a number of AGN
(see \citealt{Saikia-Salter1988:ARA&A.26}, and references
therein). They are thought to be caused by changes in the magnetic
field or shock conditions but, as we have demonstrated, may also be
due to different, unresolved components dominating the emission at
different times.

%%%%%%%%%%%%%%%%%%%%%%%%%%%%%%%%%%%%%%%%%%%%%%%%%%%
%%%%%%%%%%%%%%%%%%%%%%%%%%%%%%%%%%%%%%%%%%%%%%%%%%%
%%%%%%%%%%%%%%%%%%%%%%%%%%%%%%%%%%%%%%%%%%%%%%%%%%%

\subsection{Source distance}\label{sect:energetics}

\begin{figure}
  \centering
  \resizebox{\hsize}{!}{\includegraphics[angle=0]{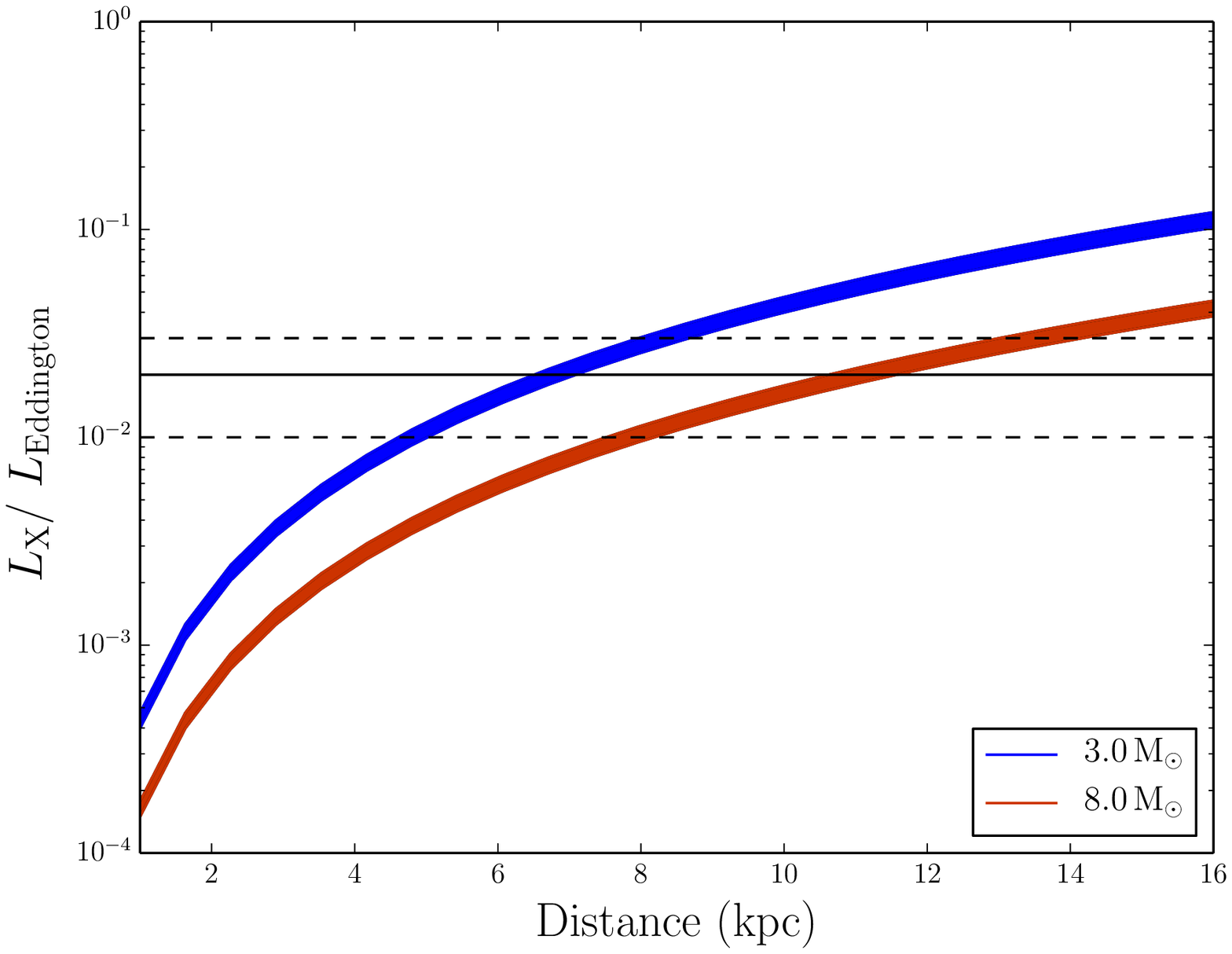}
  }
  \resizebox{\hsize}{!}{\includegraphics[angle=0]{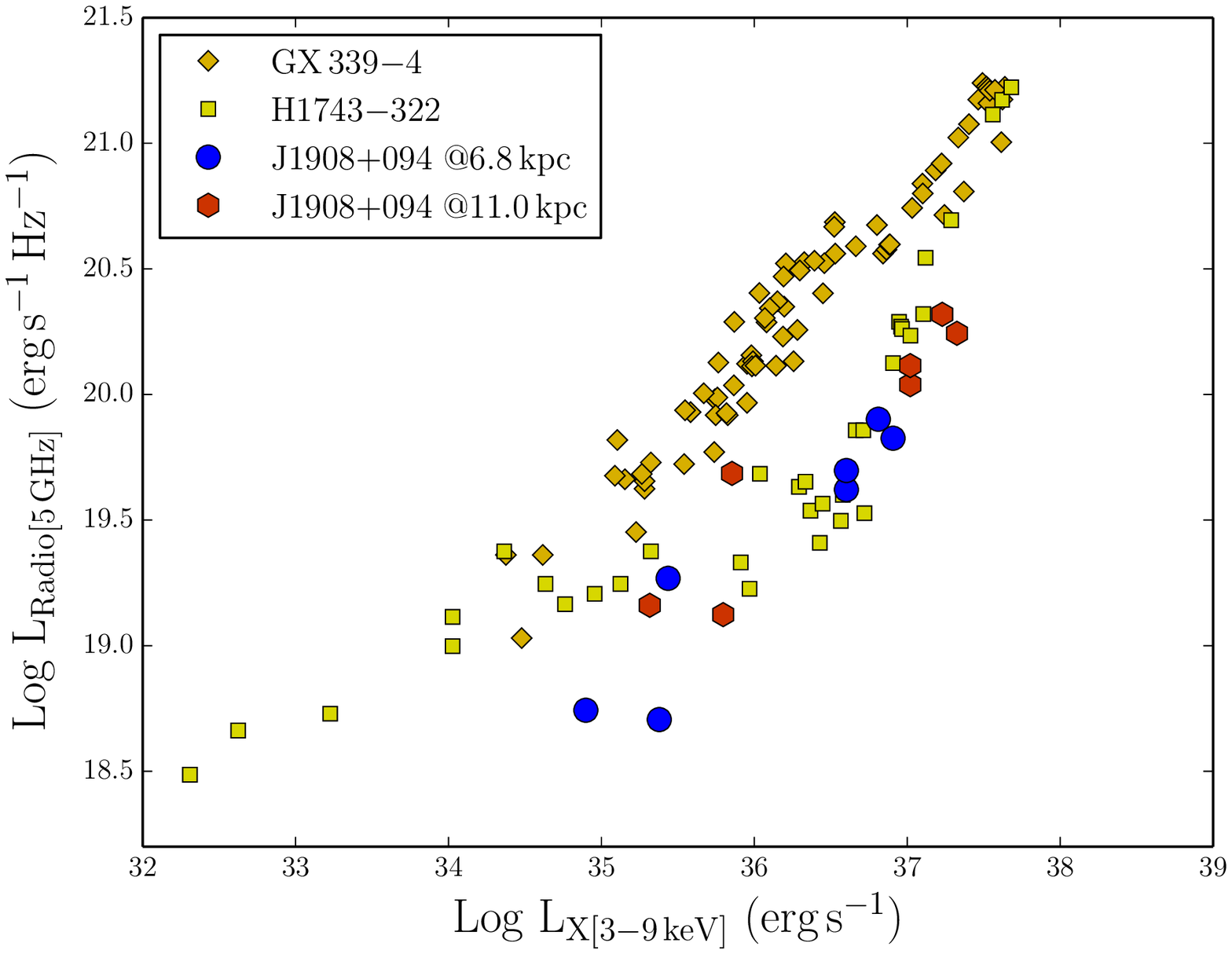}
  }
  %\resizebox{14cm}{!}{\includegraphics[angle=0]{.eps}}
  %\vspace{0mm} 
  \caption{{\it Upper panel}: Assuming that the source transitions
    from a soft to hard state at 2\% of the Eddington luminosity and a
    representative black hole mass of 8\,M$_{\odot}$
    \citep{Kreidberg2012:ApJ.757}, the observed fluxes imply a source
    distance of 11\,kpc (red), or, for a black hole mass of
    3\,M$_{\odot}$, 6.8\,kpc (blue). Width of lines corresponds to
    uncertainty in observed flux at the state transition observed in
    2002. Solid and dashed lines represent 2, 1 and 3 \% of the
    Eddington luminosity. {\it Lower panel}: The hard-state
    X-ray--radio luminosity correlation shows representative data from
    the upper (GX\,339$-$4; \citealt{Corbel2013:MNRAS.428}) and lower
    (H1743$-$322; \citealt{Coriat2011:MNRAS.414}) branches, overlaid
    by our data for \xtenineteen\ at assumed distances of 6.8\,kpc and
    11\,kpc (the latter distance being more consistent with the
    representative data).
}
  \label{fig:Luminosity} 
\end{figure}

%%% Eddington / Transition

The transition from soft to hard states at the end of X-ray binary
outbursts are observed to cluster at approximately 2\% of the
Eddington luminosity
\citep{Maccarone2003:A&A.409,Kalemci2013:ApJ.779}.  While
\cite{Dunn2010:MNRAS.403} find a significant spread of values from 0.5
to 10\%, using sources with poorly constrained masses and distances,
\cite{Kalemci2013:ApJ.779} find a much narrower spread from 1 to 3\%.
Our observations do not cover the state
transition itself but from X-ray fluxes in the soft and hard states
(on MJD 56629 and 56703) this transition must have occurred at a flux
of between $\sim$1--8 $\times 10^{-10}$ ergs\,cm$^{-2}$\,s$^{-1}$
(3--9\,keV).
This is consistent with the transition flux of 4.2--5.0 $\times
10^{-10}$ ergs\,cm$^{-2}$\,s$^{-1}$ (2.5--25\,keV) found by
\cite{Gogus2004ApJ.609} for the 2002 outburst of this source, at MJD
52425--52425.
Converting the latter values to bolometric flux as in
\cite{Maccarone2003:A&A.409}, i.e., assuming a spectrum of ${\rm
  d}N/{\rm d}E \propto E^{-1.8} \exp^{(-E/200\,{\rm keV})}$ integrated
from 0.5--10$^{4}$ keV, we find that the transition occurs at $\approx
1.3 \times 10^{-9}$ ergs\,cm$^{-2}$\,s$^{-1}$.
Assuming a representative black hole mass of 8\,M$_{\odot}$
\citep{Kreidberg2012:ApJ.757} this implies that the distance to
\xtenineteen\ is $\approx$11\,kpc (7.8--13.6 kpc using the observed
1--3\% range of transition fluxes), or $\approx$6.8\,kpc (4.8--8.3
kpc) for a black hole of 3\,M$_{\odot}$ (Figure\,
\ref{fig:Luminosity}).

%%% X-ray---Radio correlation

While there is uncertainty \citep{Gallo2014:MNRAS.445} as to whether
the correlation between the X-ray and radio luminosities of \lmxbs\ in
the hard state (e.g.,
\citealt{Hannikainen1998:A&A.337,Corbel2000:A&A.359,Corbel2003:A&A.400,Corbel2013:MNRAS.428,Gallo2003:MNRAS.344,Gallo2012:MNRAS.423})
is universal, it remains a useful diagnostic of source distance.
In Figure\,\ref{fig:Luminosity} we plot representative data from the
upper (GX\,339$-$4; \citealt{Corbel2013:MNRAS.428}, assuming, as they
do, a distance of 8\,kpc; \citealt{Zdziarski2004:MNRAS.351}) and lower
(H1743$-$322; \citealt{Coriat2011:MNRAS.414}) branches.
For comparison, we calculate the luminosities of \xtenineteen\ over/at
the same energies as \citeauthor{Corbel2013:MNRAS.428} and
\citeauthor{Coriat2011:MNRAS.414}, at both 6.8 and 11 kpc (from
above). Assuming a flat spectrum, we calculate the luminosity at
5\,GHz from both our 5.25 and 15.7 GHz observations, for comparison with
X-ray observations which occurred within 24 hours. We also plot the
8.3\,GHz radio flux on MJD 52723.5 and 52741.5 (2003) from
\citep{Jonker2004:MNRAS.351} and, using their spectral parameters, the
interpolated 3--9\,keV X-ray flux on 52722.0 and 52748.6. We did not
extrapolate these data to a common epoch as any extrapolation will be
insignificant within errors.

We find that the luminosities inferred assuming a source distance of
11\,kpc are more consistent with the representative data of other
black hole sources, though there is some scatter. Regardless
of the source distance, \xtenineteen\ appears to fall on the lower,
radio-quiet branch of the X-ray--radio flux relation
(e.g. \citealt{Coriat2011:MNRAS.414}).
The inconsistency of the 6.8\,kpc data at low X-ray luminosities is
caused by data from MJD 52741/8 (i.e., one of the 2003 points) and MJD
56727 (i.e., our late-time point); while the earlier point has a 7-day
difference between observation epochs, the later point is based on
data with only 11 hours difference, during a period with little
variability.
11\,kpc is also consistent with the lower limit of $\gtrsim$6\,kpc
estimated from the peak bolometric flux in 2002
\citep{intZand2002A&A.394}, as is 6.8\,kpc.
Though the optical extinction ($E_{B-V} \approx 5.4$;
\citealt{schlegel1998:ApJ500}) in that region ($l,b = 43.26\arcdeg,
+0.43\arcdeg$) or derived, via \citealt{Guver2009:MNRAS.400}, from our
observed X-ray extinction ($2.2 \lesssim E_{B-V} \lesssim 4.4$) is
quite high, neither are inconsistent with the detection of a
near-infrared counterpart at $\approx$11\,kpc
\citep{Chaty2002IAUC.7897,Chaty2006MNRAS.365}.

%%**************************************************
%%**************************************************
%% Section : Conclusion
%%**************************************************
%%**************************************************

\section{Conclusions}\label{section:conclusion}

We obtained multifrequency radio data of the \lmxb\ and black hole
candidate \xtenineteen\ from the VLA and AMI-LA arrays, spanning the
entirety of its 2013 outburst. We also analysed the \swift/XRT X-ray
data from this period and compared with the available MAXI and
\swift/BAT X-ray light curves.
We determine that the broadband light curves, X-ray hardness-intensity and
X-ray--radio flux relation are all consistent with \xtenineteen\ being
a black hole \lmxb.  The source traced the standard
hardness-intensity path, evolving from a hard state, through a soft
state, before returning to a hard state. The radio behaviour is
typical of a compact jet that becomes quenched and transitions to a
discrete ejection at the late stages of X-ray softening
\citep{Fender2006:csxs.book381}.
We note that there is significant radio variability on time-scales of
both hours and days and that this variability can only be observed
with daily monitoring observations -- without which many radio flares
may go undetected.
Neither the X-ray nor radio flux or light curve morphology displayed
any significant differences from the 2002 outburst of this source,
despite the obvious difference in separation of the resolved, discrete
ejecta between the two outbursts.

Radio polarisation measurements allowed us to infer the jet
orientation as being approximately in the north-northwest direction,
in agreement with the resolved observations. We also observe a sharp
90\arcdeg\ rotation of EVPA associated with a change of dominant
component in resolved observations, as well as a lower-level EVPA
drift associated with the expansion of the ejecta reported elsewhere
\citep{Rushton2014}.
Assuming that \xtenineteen\ is a 8\,M$_{\odot}$ black hole, we
estimate a distance of $\approx$11\,kpc (7.8--13.6 kpc) to the source,
based both on the bolometric X-ray flux at the soft to hard transition
and on the X-ray--radio flux relation. Regardless of the source
distance, \xtenineteen\ is the newest addition to the lower,
radio-quiet branch of the X-ray--radio flux relation (e.g.
\citealt{Coriat2011:MNRAS.414}).

%%**************************************************
%%**************************************************
\begin{table*}
  \centering	
  \caption{Radio flux densities of source, $F_{\nu}$, at frequency,
    $\nu$, and Stokes $Q$ and $U$ flux densities or 3$\sigma$ upper
    limit on polarisation, $P_{\rm{limit}}$, at 5.25 and 7.45 GHz
    (statistical errors only).  Additional 15.7\,GHz data, plotted in
    figure \ref{fig:broad_lc}, are not tabulated but have 3$\sigma$
    upper limits of $<0.24$mJy.  }
\label{table:fluxes} 
\begin{tabular}{l l l l l l} 
  \hline
  Epoch   & $\nu$  & $F_{\nu}$     & $Q$     & $U$  & $P_{\rm{limit}}$ \\
  (MJD)   & (GHz) & (mJy) & (mJy\,beam$^{-1}$) & (mJy\,beam$^{-1}$) & (mJy\,beam$^{-1}$) \\
  \hline
56594.968 & 5.25 & 0.76 $\pm$ 0.02   & ... & ...  & $<$0.07 \\ 
56594.968 & 7.45 & 0.70 $\pm$ 0.02   & ... & ...  & $<$0.06 \\ 
56595.700 & 15.7 & 0.83 $\pm$ 0.08   & ... & ...  & ...\\ 
56596.750 & 15.7 & 1.21 $\pm$ 0.06   & ... & ...  & ...\\ 
56597.700 & 15.7 & 1.62 $\pm$ 0.06   & ... & ...  & ...\\ 
56599.700 & 15.7 & 1.41 $\pm$ 0.07   & ... & ...  & ...\\ 
56600.700 & 15.7 & 1.11 $\pm$ 0.07   & ... & ...  & ...\\ 
56601.078 & 5.25 & 0.52 $\pm$ 0.02   & ... & ...  & $<$0.07 \\ 
56601.078 & 7.45 & 0.48 $\pm$ 0.03   & ... & ...  & $<$0.08 \\ 
56601.700 & 15.7 & 9.29 $\pm$ 0.09   & ... & ...  & ...\\ 
56602.700 & 15.7 & $<$0.32   & ... & ...  & ... \\ 
56603.062 & 5.25 & 1.27 $\pm$ 0.03   & 0.069 $\pm$ 0.025 & 0.103 $\pm$ 0.025  & ... \\ 
56603.062 & 7.45 & 1.14 $\pm$ 0.04   & ... & ...  & $<$0.10 \\
56603.649 & 0.145 & $<$9.3   & ... & ...  & ... \\ 
56603.700 & 15.7 & 0.67 $\pm$ 0.08   & ... & ...  & ... \\ 
56604.026 & 5.25 & 2.52 $\pm$ 0.04   & -0.034 $\pm$ 0.028 & -0.203 $\pm$ 0.028  & ... \\ 
56604.026 & 7.45 & 2.33 $\pm$ 0.04   & -0.058 $\pm$ 0.032 & 0.194 $\pm$ 0.031  & ... \\ 
56604.700 & 15.7 & 3.03 $\pm$ 0.11   & ... & ...  & ...\\ 
56604.958 & 22 & 2.45 $\pm$ 0.03   & ... & ...  & ...\\ 
56605.076 & 5.25 & 4.73 $\pm$ 0.04   & -0.380 $\pm$ 0.026 & -0.341 $\pm$ 0.026  & ... \\ 
56605.076 & 7.45 & 4.00 $\pm$ 0.04   & 0.153 $\pm$ 0.031 & 0.521 $\pm$ 0.030  & ... \\ 
56605.700 & 15.7 & 10.40 $\pm$ 0.10   & ... & ...  & ... \\ 
56606.700 & 15.7 & 12.91 $\pm$ 0.07   & ... & ...  & ... \\ 
56606.918 & 22 & 11.47 $\pm$ 0.04   & ... & ...  & ... \\ 
56607.070 & 5.25 & 13.43 $\pm$ 0.11   & -0.120 $\pm$ 0.026 & -0.091 $\pm$ 0.025  & ... \\ 
56607.070 & 7.45 & 12.96 $\pm$ 0.10   & 0.158 $\pm$ 0.035 & 0.289 $\pm$ 0.035  & ... \\
56607.700 & 15.7 & 4.63 $\pm$ 0.12   & ... & ...  & ...\\ 
56608.050 & 22 & 4.71 $\pm$ 0.03   & ... & ...  & ...\\ 
56608.098 & 5.25 & 6.72 $\pm$ 0.07   & 0.237 $\pm$ 0.028 & 0.287 $\pm$ 0.029  & ... \\ 
56608.098 & 7.45 & 6.14 $\pm$ 0.06   & 0.045 $\pm$ 0.033 & -0.320 $\pm$ 0.033  & ... \\ 
56608.635 & 0.145 & $<$12.5   & ... & ...  & ... \\ 
56608.700 & 15.7 & 2.74 $\pm$ 0.08   & ... & ...  & ...\\ 
56609.048 & 22 & 2.32 $\pm$ 0.02   & ... & ...  & ...\\ 
56609.065 & 5.25 & 3.33 $\pm$ 0.04   & 0.133 $\pm$ 0.025 & 0.098 $\pm$ 0.026  & ... \\ 
56609.065 & 7.45 & 3.05 $\pm$ 0.04   & -0.087 $\pm$ 0.029 & -0.129 $\pm$ 0.028  & ... \\ 
56609.700 & 15.7 & 1.24 $\pm$ 0.08   & ... & ...  & ...\\ 
56610.045 & 22 & 1.27 $\pm$ 0.03   & ... & ...  & ...\\ 
56610.062 & 5.25 & 2.22 $\pm$ 0.03   & ... & ...  & $<$0.08 \\ 
56610.062 & 7.45 & 2.04 $\pm$ 0.03   & ... & ...  & $<$0.08 \\ 
56610.700 & 15.7 & 0.51 $\pm$ 0.07   & ... & ...  & ...\\ 
56611.042 & 22 & 0.58 $\pm$ 0.02   & ... & ...  & ...\\ 
56611.601 & 0.145 & $<$12.6   & ... & ...  & ... \\ 
56611.700 & 15.7 & 0.49 $\pm$ 0.07   & ... & ...  & ...\\ 
56612.040 & 1.6 & 2.09 $\pm$ 0.24   & ... & ...  & ...\\ 
56612.078 & 5.25 & 0.98 $\pm$ 0.06   & ... & ...  & $<$0.17 \\ 
56612.078 & 7.45 & 0.91 $\pm$ 0.07   & ... & ...  & $<$0.21 \\ 
56614.597 & 0.145 & $<$13.5   & ... & ...  & ... \\ 
56614.700 & 15.7 & $<$0.30   & ... & ...  & ... \\ 
56615.700 & 15.7 & $<$0.24   & ... & ...  & ... \\ 
56616.004 & 5.25 & 0.38 $\pm$ 0.05   & ... & ...  & $<$0.14 \\ 
56616.004 & 7.45 & 0.29 $\pm$ 0.05   & ... & ...  & $<$0.15 \\ 
56616.008 & 1.6 & 0.92 $\pm$ 0.08   & ... & ...  & ...\\ 
56616.029 & 22 & 0.14 $\pm$ 0.03   & ... & ...  & ...\\ 
56616.700 & 15.7 & $<$0.43   & ... & ...  & ... \\ 
56617.700 & 15.7 & $<$0.24   & ... & ...  & ... \\ 
56617.999 & 5.25 & 0.36 $\pm$ 0.03   & ... & ...  & $<$0.07 \\ 
56617.999 & 7.45 & 0.27 $\pm$ 0.03   & ... & ...  & $<$0.09 \\ 
56618.020 & 1.6 & 0.53 $\pm$ 0.03   & ... & ...  & ...\\ 
56618.700 & 15.7 & $<$0.24   & ... & ...  & ... \\ 
56727.532 & 5.25 & 0.09 $\pm$ 0.02   & ... & ...  & $<$0.05 \\ 
56727.533 & 7.45 & 0.09 $\pm$ 0.02   & ... & ...  & $<$0.05 \\ 
56738.694 & 5.25 & $<$0.07   & ... & ...  & ... \\ 
56738.694 & 7.45 & $<$0.09   & ... & ...  & ... \\ 
\hline
\end{tabular}
\end{table*}
%%**************************************************
%%**************************************************

%%**************************************************
%% Acknowledgements
%%**************************************************

\section*{Acknowledgements}

We thank the  anonymous referee for constructive comments. 
We thank T.J. Maccarone for useful discussions and A.M.M. Scaife for
valuable input on the AMI-LA data.
This work was supported by Australian Research Council grant
DP120102393.
DA acknowledges support from the Royal Society.
GRS is supported in part by an NSERC Discovery Grant.  
SM is supported by the Spanish Ministerio de Ciencia e Innovaci\'on
(S.M.; grant AYA2013-47447-C03-1-P).
This work is supported in part by European Research Council Advanced
Grant 267697 `4 Pi Sky: Extreme Astrophysics with Revolutionary Radio
Telescopes'.
The National Radio Astronomy Observatory is a facility of the National
Science Foundation operated under cooperative agreement by Associated
Universities, Inc.
We thank the staff of the Mullard Radio Astronomy Observatory for
their invaluable assistance in the operation of AMI-LA.
LOFAR, the Low Frequency Array designed and constructed by ASTRON, has
facilities in several countries, that are owned by various parties
(each with their own funding sources), and that are collectively
operated by the International LOFAR Telescope (ILT) foundation under a
joint scientific policy.
This research has made use of NASA's Astrophysics Data System. 
Swift XRT data was supplied by the UK Swift Science Data Centre at the
University of Leicester and Swift BAT transient monitor results were
provided by the Swift/BAT team.

\begin{table}
  \centering	
  \caption{\swift\ observation ID, $ObsID$, XRT X-ray fluxes of source
    from 3--9\,keV, $F_{{\rm X[3-9\,keV]}}$, and the photon index of
    the power-law fit to spectra, $\Gamma$.}
\label{table:fluxes_X} 
\begin{tabular}{l l l l} 
  \hline
  Epoch   & $ObsID$  & $F_{{\rm X[3-9\,keV]}}$                   & $\Gamma$    \\
  (MJD)   &        & ($\times 10^{-10}$ &  \\
          &        & ergs\,cm$^{-2}$\,s$^{-1}$) &  \\

  \hline
%
%56729.08 & 00033014038 & 0.37 $\pm$ 0.07  & 1.48 $\pm$ 0.33 \\ 
%
56594.84 & 33014001 &	  7.23 $\pm$ 0.19  	& 1.80 $\pm$ 0.07 \\ 
56597.91 & 33014002 &	  11.70 $\pm$ 0.20  	& 2.12 $\pm$ 0.05 \\ 
56599.53 & 33014003 &	  14.60 $\pm$ 0.23  	& 2.84 $\pm$ 0.05 \\ 
56604.85 & 33014004 &	  13.80 $\pm$ 0.23  	& 4.71 $\pm$ 0.05 \\ 
56605.80 & 33014005 &	  13.20 $\pm$ 0.23  	& 4.73 $\pm$ 0.05 \\ 
56606.05 & 33014006 &	  15.10 $\pm$ 0.34  	& 4.61 $\pm$ 0.08 \\ 
56607.39 & 33014007 &	  14.50 $\pm$ 0.24  	& 4.44 $\pm$ 0.06 \\ 
56608.05 & 33014008 &	  13.40 $\pm$ 0.21  	& 4.49 $\pm$ 0.06 \\ 
56609.19 & 33014009 &	  10.20 $\pm$ 0.17  	& 4.91 $\pm$ 0.06 \\ 
56610.79 & 33014010 &	  11.40 $\pm$ 0.24  	& 4.97 $\pm$ 0.07 \\ 
56611.33 & 33014011 &	  7.57 $\pm$ 0.17  	& 5.08 $\pm$ 0.07 \\ 
56612.60 & 33014012 &	  11.70 $\pm$ 0.21  	& 4.97 $\pm$ 0.06 \\ 
56614.40 & 33014014 &	  9.18 $\pm$ 0.19  	& 5.04 $\pm$ 0.07 \\ 
56615.73 & 33014015 &	  8.74 $\pm$ 0.18  	& 4.86 $\pm$ 0.07 \\ 
56616.13 & 33014016 &	  8.88 $\pm$ 0.18  	& 4.95 $\pm$ 0.07 \\ 
56619.06 & 33014017 &	  8.06 $\pm$ 0.18  	& 5.03 $\pm$ 0.07 \\ 
56624.47 & 33014018 &	  7.39 $\pm$ 0.17  	& 4.86 $\pm$ 0.08 \\ 
56629.00 & 33014019 &	  7.79 $\pm$ 0.16  	& 5.11 $\pm$ 0.07 \\ 
56703.93 & 33014020 &	  1.13 $\pm$ 0.09  	& 2.36 $\pm$ 0.18 \\ 
56704.67 & 33014021 &	  1.07 $\pm$ 0.06  	& 2.44 $\pm$ 0.15 \\ 
56705.87 & 33014022 &	  1.09 $\pm$ 0.06  	& 1.84 $\pm$ 0.15 \\ 
56706.67 & 33014023 &	  1.08 $\pm$ 0.05  	& 2.11 $\pm$ 0.12 \\ 
56708.67 & 33014024 &	  1.02 $\pm$ 0.02  	& 2.13 $\pm$ 0.17 \\ 
56713.00 & 33014025 &	  0.83 $\pm$ 0.11  	& 2.23 $\pm$ 0.25 \\ 
56713.13 & 33014026 &	  1.28 $\pm$ 0.06  	& 1.67 $\pm$ 0.23 \\ 
56713.20 & 33014027 &	  0.64 $\pm$ 0.01  	& 1.52 $\pm$ 0.23 \\ 
56715.13 & 33014028 &	  0.55 $\pm$ 0.05  	& 1.79 $\pm$ 0.22 \\ 
56717.46 & 33014029 &	  0.54 $\pm$ 0.07  	& 1.82 $\pm$ 0.23 \\ 
56718.74 & 33014030 &	  0.68 $\pm$ 0.12  	& 2.03 $\pm$ 0.28 \\ 
56719.33 & 33014031 &	  0.65 $\pm$ 0.10  	& 1.72 $\pm$ 0.27 \\ 
56721.72 & 33014032 &	  0.53 $\pm$ 0.06  	& 1.41 $\pm$ 0.24 \\ 
56723.73 & 33014033 &	  0.66 $\pm$ 0.10  	& 1.99 $\pm$ 0.28 \\ 
56723.19 & 33014034 &	  0.57 $\pm$ 0.08  	& 1.46 $\pm$ 0.24 \\ 
56725.93 & 33014035 &	  0.47 $\pm$ 0.08  	& 1.89 $\pm$ 0.28 \\ 
56727.06 & 33014036 &	  0.43 $\pm$ 0.08  	& 1.94 $\pm$ 0.32 \\ 
56729.08 & 33014038 &	  0.38 $\pm$ 0.07  	& 1.43 $\pm$ 0.32 \\ 
\hline
\end{tabular}
\end{table}

%\vspace{-6mm}

\label{lastpage}

\end{document}